# AGGRESSION AND "HATE SPEECH" IN COMMUNICATION OF MEDIA USERS: ANALYSIS OF CONTROL CAPABILITIES[1]

V.V. Kazhberova[*], A.G. Chkhartishvili[**a], D.A. Gubanov[**b], I.V. Kozitsin[**c], E.V. Belyavsky[**d], D.N. Fedyanin[**e], S.N. Cherkasov[**f], D.O. Meshkov[**g]

[*] M.V. Lomonosov Moscow State University, Moscow, Russia, Russia, e-mail: kazhberovavv@gmail.com

[**] V.A. Trapeznikov Institute of Control Sciences Russian Academy of Sciences, Moscow, Russia, e-mail:
[a] sandro_ch@mail.ru, [b] dmitry.a.g@gmail.com, [c] kozitsin.ivan@mail.ru, [d] evgenetc@gmail.com,
[e] dfedyanin@inbox.ru, [f] cherkasovsn@mail.ru, [g] dmitrymeshkov@mail.ru

**Abstract**. Analyzing the possibilities of mutual influence of users in new media, the researchers found a high level of aggression and hate speech when discussing an urgent social problem - measures for COVID-19' fighting. This fact determined the central aspect of the research at the next stage and the central topic of the proposed article. The first chapter of the article is devoted to the characteristics of the prerequisites of the undertaken research, its main features. The following chapters include methodological features of the study, theoretical substantiation of the concepts of "aggression" and "hate speech", identification of systemic connections of these concepts with other characteristics of messages. The result was the creating of a mathematical aggression growth model and the analysis of its manageability using basic social media strategies. The results can be useful for developing media content in a modern digital environment.

[1] This work has been submitted to Journal «Vestnik Moskovskogo universiteta. Seriya 10. Zhurnalistika» for possible publication. This work was partially supported by the Russian Foundation for Basic Research, project no. 20-04-60296 (Gubanov D.A., Kozitsin I.V.).

# АГРЕССИЯ И «ЯЗЫК ВРАЖДЫ» В ОБЩЕНИИ МЕДИАПОЛЬЗОВАТЕЛЕЙ: АНАЛИЗ ВОЗМОЖНОСТЕЙ УПРАВЛЕНИЯ[2]

В.В. Кажберова[*], А.Г. Чхартишвили[**,a], Д.А. Губанов[**,b], И.В. Козицин[**,c], Е.В. Белявский[**,d], Д.Н. Федянин[**,e], С.Н. Черкасов[**,f], Мешков Д.О.[**,g]

[*] *Московский государственный университет имени М.В. Ломоносова (г. Москва, Россия), e-mail: kazhberovavv@gmail.com*

[**] *Институт проблем управления им. В.А. Трапезникова РАН (г. Москва, Россия), e-mail:*
[a] sandro_ch@mail.ru, [b] dmitry.a.g@gmail.com, [c] kozitsin.ivan@mail.ru, [d] evgenetc@gmail.com, [e] dfedyanin@inbox.ru, [f] cherkasovsn@mail.ru, [g] dmitrymeshkov@mail.ru

**Аннотация**. Статья основана на данных многоаспектного междисциплинарного исследования актуальной проблемы современности – возможностей управления экстремальными ситуациями в конкретных условиях. Цель исследования была продиктована необходимостью выявить уровень взаимного влияния Интернет-пользователей на их поведение по отношению к вакцинации во время пандемии, вызванной COVID-19. В ходе исследования обнаружилось, что взаимодействие пользователей в паблике одного из ведущих СМИ характеризует высокий уровень агрессии и «языка вражды» (в 53% комментариев). Была создана математическая модель распространения агрессии в сети и проведен анализ управляемости ее уровнем в рамках традиционных стратегий: «жестких» (сокращение числа агрессивных комментариев) и «мягких» (добавление нейтральных или позитивно окрашенных сообщений). Оказалось, что даже применение этих стратегий не позволяет снизить уровень агрессии ниже определенного значения.

**Ключевые слова**: агрессия, язык вражды, новые медиа, социальные сети, пользователи сети, комментарии, медиапотребители.

## Предпосылки и актуальность исследования

Сегодня человечество переживает эпоху четвертой промышленной революции, когда главным ресурсом становятся информация и знания, компетенции, которые можно назвать также цифровым капиталом (Швабб, 2018: 12)[1]; технологии привнесли в мир

---

медиа «разнообразие форм и доступность потенциальных источников, свободу от большинства видов контроля, интерактивность» (МакКуэйл, 2013:26).

В настоящее время в России более 90% потребителей в возрасте до 55 лет пользуются Интернетом. 76% пользователей при этом пользуется смартфонами (мобильный Интернет), 62% - десктопами, 22% - Смарт ТВ, 20% - планшетами [2]. Лидируют среди программ и приложений социальные медиа, мобильное видео и ТВ, мессенджеры, банки, Интернет-магазины.

Среди пользователей социальными медиа 72,7% россиян. Это люди разного возраста: активный прирост происходит в сегменте аудитории 50+. Средний медиапотребитель проводит в социальных сетях 2 часа 27 минут[3].

Влияние социальных медиа на человека изучается практически с самого начала их широкого распространения – с 2000-х годов. Итоги исследования, продолжавшегося с 2003 по 2018 год, показывают, что наиболее травмирующими эмоциями пользователей соцсетей стали ощущение паники, раздражение, стресс, тревога, депрессия, чувства вины и одиночества (Boroon, Abedin, Erfani, 2021: 1-21).

В России идентичных по длительности наблюдения и анализа исследований не проводилось. Однако периодические «срезы» отношения пользователей в целом подтверждают общую тенденцию.

Например, отмечается, что в России достаточно высок уровень киберагрессии[4]: по данным исследования Microsoft DCI, 71% пользователей отметили, что были оскорблены в Интернете (в мире этот показатель ниже на 20%), при этом 60% россиян получили негативный комментарий от совершенно незнакомых людей (в мире этот показатель на уровне 36%)[5]. Согласно данным другого исследования, 49% респондентов испытывают негативные эмоции, пользуясь социальными медиа: 34% пользователей признались в зависимости от онлайн-платформ, 26% сообщили, что были расстроены агрессивными сообщениями в свой адрес[6]. Эти выводы дополнили качественные исследования, обозначившие риски обезличивания, анонимности, асоциальности, потери приватности, мошенничества при общении в сети, что может травмировать человека (Ефимова, Зюбан, 2016; Аптикиева, 2020).

Нельзя утверждать, что причины этих «теневых» сторон соцсетей заключены только в особенностях медиаплатформы. Исследователи разных областей знания констатируют, что в обществе происходит пересмотр ценностей, социальных ролей, вызванный как манифестацией традиций постмодернизма, так и феноменом, который был назван *эрой постправды*. Он тесно связан с распространением фейков, пранком и троллингом – все эти явления объединяет сознательный отказ от достоверности, грубое вмешательство в информационное пространство медиапотребителя (Зиновьев, 2018; Воронцова, 2016: 109-115). Фейки, например, американские пользователи признали более опасным явлением, чем расизм и терроризм[7].

Чтобы исследовать и систематизировать процесс взаимного влияния медиапотребителей, в июле 2021 года авторами статьи был запущен научный проект по изучению процессов формирования и изменения представлений пользователей онлайновых социальных сетей. Тема мер против COVID-19, в частности, вакцинации населения, была выбрана не случайно: она оказалась социально значимой и потенциально конфликтной, поскольку сильно зависит от взглядов и ценностей человека, а также отношений в социуме (Бузина, Бузин, Ланской, 2020). Даже ношение медицинских масок в период пандемии сильно коррелирует с политическими взглядами россиян (Chkhartishvili, Gubanov, Kozitsin, 2021).

В ходе изучения взаимодействия пользователей оказалось, что значительная часть материалов содержит агрессию и «язык вражды»: они встретились в каждом втором комментарии. Это намного выше обычных значений: по данным *BrandAnalytics*, доля агрессии в социальных медиа в среднем составляет примерно 5,5% от всех сообщений[8].

Основной целью исследования стало выявление системных взаимосвязей, позволяющих описать актуальную модель агрессии в социальных медиа и возможности управления ею.

**Теоретическая база исследования и смежные направления**

Сама идея изучения взаимовлияния медиапотребителей основана на исследованиях того, как социальное влияние может изменять представления и действия людей – это модели Френча-Харари-ДеГроота, Фридкина-Джонсона и др. В рамках этого подхода выделяются макро- и микромодели, а также разные виды влияния (Губанов, Новиков, Чхартишвили, 2018).

Известно, что мнения пользователей социальных медиа могут быть идентифицированы с помощью доступных для наблюдения действий («цифровых отпечатков»): постов, лайков, подписок, комментариев. Методы их изучения позволяют наблюдать поведение человека долговременно, в естественной для него коммуникационной среде (в отличие, например, от опросов). Однако практических исследований такого влияния пока недостаточно, что еще раз подчеркивает их актуальность.

Социальные сети – часть медиасистемы, которая развивается согласно своим внутренним закономерностям, поэтому авторы статьи учитывали также данные актуальных исследований, описывающих роль современных медиа в освещении конфликтов (Смирнова, 2021; Демина, Шкондин, 2021), а также посвященных особенностям новой реальности, сформированной пандемией коронавируса (Смирнова, Шкондин, Денисова, Стебловская, 2022).

В отношении агрессии и «языка вражды» в целом на протяжении последних 25 лет сложился устойчивый научный дискурс (Сковородников, 1997; Быкова, 1999; Дзялошинский, 2006; Дзялошинский, 2019). В его рамках авторы статьи опирались на действующие определения агрессии и «языка вражды», а также границы применения этих понятий:

- *«язык вражды»* – это выражение дискриминации одной группы – этнической, расовой, национальной, религиозной, языковой; содержит противопоставление «мы-они», нетолерантные высказывания;

- *агрессия* – специфическая форма речевого поведения (в том числе, в письменной речи), которая мотивирована агрессивным состоянием говорящего.

В исследованиях агрессии и «языка вражды» внимание уделяется, в основном, традиционным СМИ. Наблюдается дефицит актуальных работ в контексте новых медиа, в то время как ситуация меняется очень динамично.

Ряд исследований появляется на стыке наук, например, медиакоммуникаций и психологии, социологии. Выявлено, например, что агрессия в сети может быть обусловлена несправедливыми действиями государства, материальным положением пользователя, его коммуникативными навыками, навязанными социумом социальными ролями (Рябов, Боченкова, 2021: 170-178). Некоторые исследования изучают агрессию в соцсетях как часть сетевого насилия в целом: буллинга, хейтинга, призывов к убийству и самоубийству и т.д. (Малахаева, Потапов, 2018: 724-740; Солдатова, Рассказова, Чигарькова, 2020: 3-20).

На фоне таких тенденций все острее становится вопрос о возможности минимизации подобных проявлений. Так, в последние годы трансформируется представление о возможностях управления социальными системами и, в частности, социальными медиа (Губанов, Новиков, Чхартишвили, 2018).

В целом, в мировой практике растет интерес к использованию социальных медиа в качестве инструментов для изменения поведенческих характеристик в определенных целях. В этом контексте используется термин «сетевых вмешательств», от англ. *network interventions* (Valente, 2012: 49-53; Kempe, Kleinberg, Tardos, 2003: 137-146). Они могут быть индивидуальными, групповыми, индукционными (т.е. вызывать целую серию поведенческих изменений, называемую «каскадом»), структурными. Часто их влияние ограничено определенной сферой применения, например, здравоохранением (Cobb, Josee, 2014; Hunter, 2019; Kim, Hwong, Stafford, 2015). Но, так или иначе, пока технологические, мировоззренческие и этические особенности применения этих технологий в российских социальных медиа не вполне понятны.

В настоящий момент очевидно, что изучение агрессии и «языка вражды» в социальных медиа может дать представление не только о влиянии пользователей друг на друга, но и о самом характере горизонтальных (по принципу «равный – равному») коммуникаций в современном социуме. В этом контексте исследования, изучающие риски такого взаимодействия, очень актуальны.

**Методика исследования**

В качестве эмпирической базы исследования были использованы материалы паблика крупного российского СМИ – «РосБизнесКонсалтинг» (далее – РБК) на платформе онлайновой социальной сети «ВКонтакте» [9], где пользователи активно обсуждали вопросы, связанные с пандемией COVID-19 и ее последствиями, включая массовую вакцинацию.

Для анализа автоматизированными методами были выбраны 5 тыс. постов по тематике COVID-19, а также комментарии пользователей на эти посты за период с 19 января 2020 года по 30 июня 2021 года (390 тыс. комментариев и 970 тыс. лайков). Затем из этого массива по ключевым словам: «вакцина», «вакцинация», «Спутник V», «ЭпиВакКорона», «КовиВак» было отобрано 17 тыс. комментариев, релевантных теме вакцин и вакцинации. Для анализа использовались те из них, у которых были исходные, «родительские» сообщения (в ответ на которые размещался комментарий).

<u>*На первом этапе исследования*</u> экспертами проекта (В.В. Кажберова, С.Н. Черкасов, Д.О. Мешков) был вручную обработан массив из 4,8 тыс. комментариев: эксперты маркировали комментарии, определяя их отношение к вакцинации (да/нет/непонятно), а также предполагаемое влияние содержания комментария на других пользователей (положительное; отрицательное; нейтральное/непонятное).

Значительная часть комментариев при этом не подлежала анализу, поскольку содержала сарказм, оскорбления, обесценивание чужих чувств и эмоций – это затрудняло определение реальной позиции пользователя. С точки зрения влияния, такие сообщения, согласно рабочей гипотезе экспертов проекта, негативно влияли на убеждения других пользователей, создавая деструктивный, потенциально травмирующий контекст обсуждения.

Таким образом, обозначилась необходимость более детального изучения комментариев, содержащих «язык вражды» и агрессию.

Для *второго этапа исследования* из уже обработанной выборки был случайным образом выделен массив данных из 400 комментариев, релевантных теме вакцин и вакцинации, а также соответствующих им «родительских» сообщений (еще 220 комментариев). Для их вторичного анализа экспертами был разработан ***следующий классификатор***:

- *отношение к вакцинации* (0 – против вакцин и/или вакцинации, 1 – за вакцины и/или вакцинацию, 2 – непонятная/ нейтральная позиция);

- *наличие логического обоснования своей позиции* (1 – наличие логики, 0 – ее отсутствие);

- *упоминание личного опыта (позитивного, негативного) или опыта ближайшего окружения* (1 – наличие упоминания, 0 – отсутствие);

- *наличие «языка вражды»* (1 – наличие, 0 – отсутствие);

- *наличие агрессии по отношению к оппоненту/собеседнику* (1 – наличие, 0 – отсутствие);

- *наличие агрессии по отношению к другим субъектам или объектам* (1 – наличие, 0 – отсутствие).

Поле *наличие логического обоснования своей позиции* и *упоминание личного опыта* было введено, поскольку у авторов исследования появилась гипотеза, что комментаторы, использующие такого рода аргументацию, менее эмоциональны и поэтому менее склонны к агрессии и использованию «языка вражды». По этому параметру оценивалось, использует ли человек какую-либо аргументацию, приводит ли ссылки на публикации, в том числе научные.

Поле *наличие агрессии по отношению к другим субъектам или объектам* появилось как уточняющее, помогающее разделить сообщения с агрессией к собеседнику и сообщения, где объект вражды носит более абстрактный характер (страны, вакцины, политические акторы, компании и т.д.).

Таким же образом маркировались «родительские» комментарии. Для обозначения характеристики родительского комментария применялся индекс *p*, например, «мнение$_p$».

В сложных случаях решение принималось на основе совместного экспертного обсуждения.

Для анализа комментариев, размеченных вручную экспертами, применялись методы структурирования и обработки данных (библиотека Python Data Analysis Library), методы численных вычислений (библиотека Numerical Python) и методы для выполнения научных и инженерных расчётов (библиотека Scientific Library for Python)[10].

### Результаты исследования

*Итоги контент-анализа*

По итогам <u>первого этапа исследования</u> – экспертного анализа 4,8 тыс. комментариев была выявлена структура отношения пользователей к вакцинации. Значительная часть из них связывала эту тему с политикой и острой социальной проблематикой, где комментаторы не всегда могли удержаться в морально-этических рамках.

Доля сообщений «за» вакцинацию оказалась относительно небольшой (11%). В качестве аргументации пользователи часто приводили личный опыт или опыт близких (*«Родители боялись противопоказаний, но сделали прививку. В итоге все хорошо…»*), рассуждали об общей логике развития пандемии (*«Коронавирус не закончится, если не привьемся»*), или же ограничивались оценочными суждениями (*«Спутник лучший!»*). Наряду с признанием пользы вакцинации встречались грубые оскорбления оппонентов: чаще всего, упреки в непонимании ситуации, ангажированной позиции. Такие комментарии чаще всего маркировались как поддерживающие вакцинацию (графа «мнение»), но оказывающие негативное влияние на других пользователей (графа «влияние»).

В существенно большей доле сообщений была высказана позиция «против» вакцинации (24%). В этих комментариях было отмечено большое количество мифов, бездоказательных утверждений (*"вирус мутирует граждане, и делать прививку бесполезно!"; "не наводите панику! это обычное ОРВИ"; "вакцины придумали, чтобы сократить население земли"; "после вакцины вы умрете еще быстрее!"; "мы против опытов над людьми! А как же Нюрнбергский процесс?"; "Есть VIP-вакцина более высокой степени очистки, ею привилась элита"*). Среди мифологем встречались также связи между появлением COVID-19 и внедрением технологий 5G, а также планами Билла Гейтса, Илона Маска и т.д., однако в структуре комментариев они не носили массовый характер.

Противники вакцинации также часто были агрессивны: желали оппонентам болезни или смерти, размещали уничижительные, саркастические, оскорбительные замечания.

Примеры оскорбительных комментариев (орфография и пунктуация сохранены):
- *"Тебя азербота от бешенства будут вакцинировать"*
- *"Пусть врачи и учителя покажут пример вакцинации…"*

- *"Можно подумать русский продристанный васёк из иркутстка что-то знает про западных учёных, вузы и лаборатории, а не хлебает помои на полоумный скот с раши тудей"*

- *"Ну, вот ты от естественного отбора и подохнешь. Что ты привился - молодец, в конце-концов, людям нужно испытать вакцину на биомусор... на добровольцах"*

- *"Вот бот фейковый - мозгов смотрю тебе для беседы не подвезли, оскорблять начала.. Сама иди и что-то там отрабатывай, тупая курица"*.

Больше всего оказалось сообщений, где позиция пользователя была маркирована как непонятная (нейтральная). Их доля составила 65%.

Некоторые пользователи выражали вполне обоснованное отношение к вакцинальной кампании, отмечая ее недостатки, например (орфография и пунктуация сохранены):

- *"вакцина политизирована… не участвуй политики, мы бы привились"*

- *"Нужно больше реальных мнений экспертов. Сейчас мнения слишком полярны, выглядят заказными с обеих сторон"*

- *"не нужно было ругать западные вакцины, это подорвало идею вакцинации в целом"*

- *"работа с населением не качественная. И тут еще и речи про принудиловку только усугубляют ситуацию, ибо много параноиков развелось"*.

Таким образом, контент-анализ выявил поляризацию мнений, их значительную политизированность, а также высокий уровень агрессии и «языка вражды».

***Анализ распределения комментариев по основным критериям и взаимосвязь между ними***

В рамках *второго этапа исследования* – анализа 620 комментариев было проанализировано их распределение по классификатору, описанному в части «Методика исследования». Данные представлены на рисунке (см. Рис. 1).

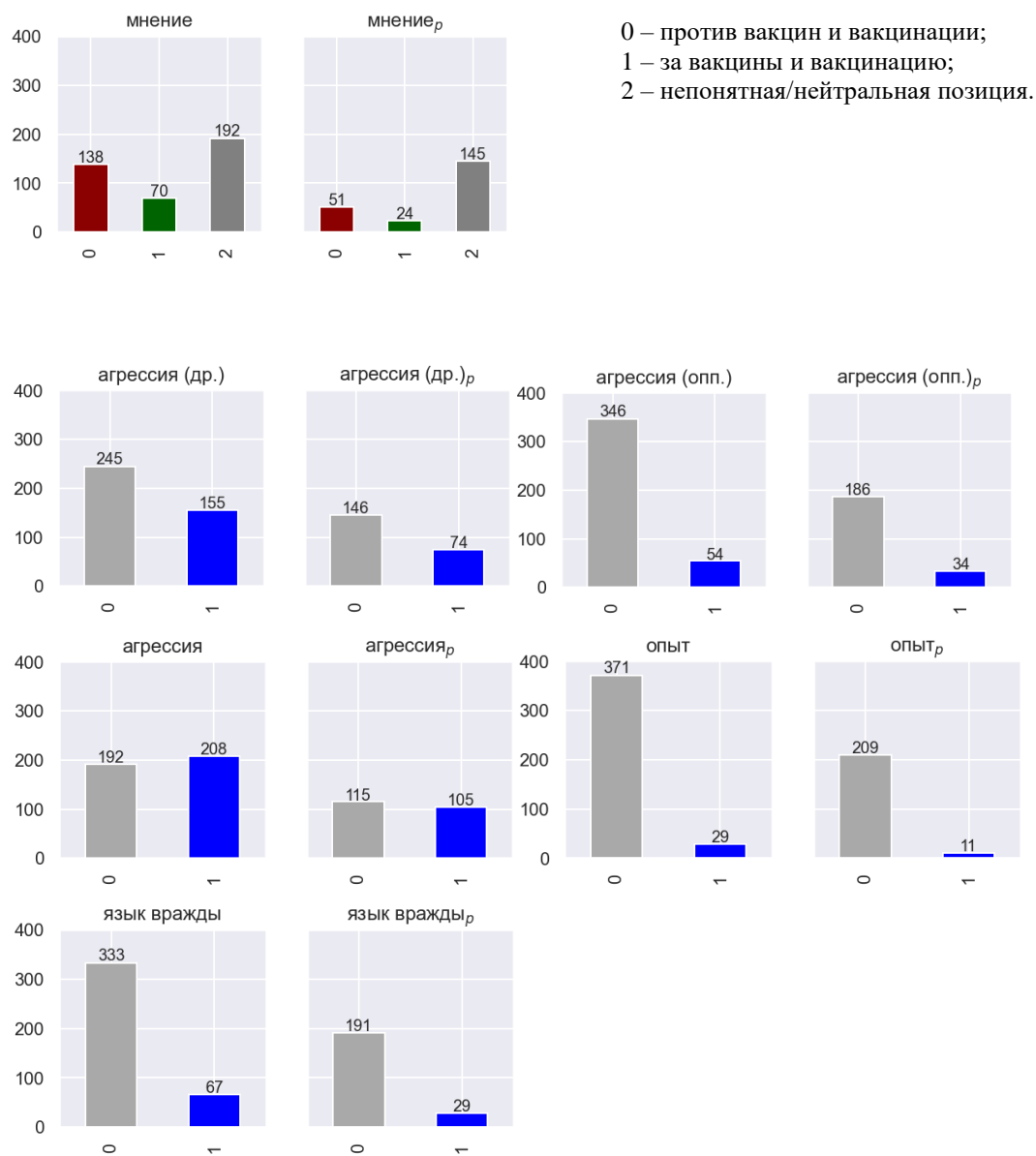

*Рис. 1.* **Распределение комментариев по классификатору**

Где *агрессия (опп.)* – агрессия к оппоненту;
*агрессия (др.)* – агрессия к другим объектам;
*значение$_р$* – значение «родительского» комментария.
**0 – отсутствие признака; 1 – наличие признака.**

Как показал анализ, пользователи в своих высказываниях редко прибегали к логическому обоснованию своей позиции (всего в 4% случаев), немного чаще ссылались на личный опыт или опыт окружения (7%). Довольно много выражений агрессии в отношении оппонента (14%), много – в отношении других объектов и субъектов (39%). Общий уровень агрессии составил 53%. Доля сообщений с «языком вражды» - 17%.

Дальнейший анализ осуществлялся при помощи факторных таблиц, которые показали совместное распределение признаков (переменных), разделенных попарно. Обобщенные результаты представлены в виде «тепловой карты» (см. Рис. 2).

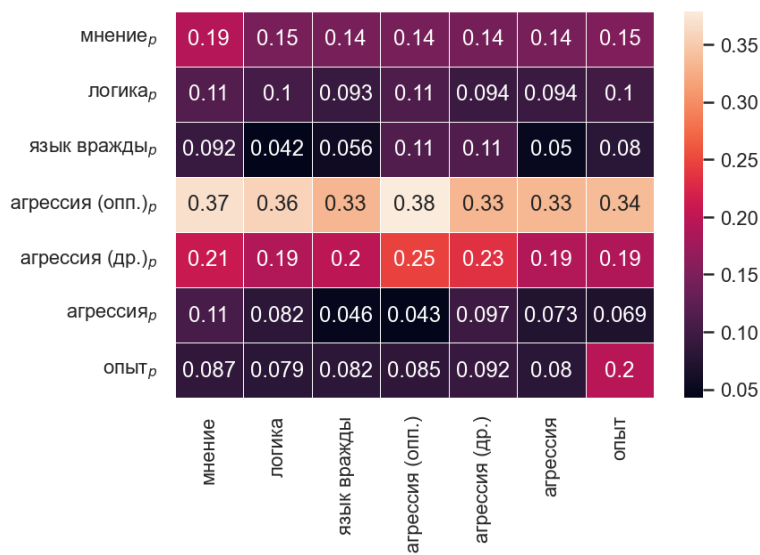

*Рис. 2*. **Итоги анализа силы связей между характеристиками классификатора**

Не все характеристики классификатора показали сильные взаимосвязи между собой. Например, по параметру *отношение к вакцинации* явной детерминированности выявлено не было, что говорит о том, что мнения в ветках дискуссий чередуются: охотнее на пост/комментарий отвечают те, кто с ним не согласен, чем те, которые поддерживают высказанные в нем идеи.

Более других оказались связаны между собой такие параметры, как *отношение к вакцинации* и *наличие агрессии*, *отношение к вакцинации* и *наличие «языка вражды»*, а также *отношение к вакцинации* и *наличие логики*.

Так, для *позиции «против» вакцинации* характерен высокий уровень агрессии по отношению к другим объектам, уровень агрессии в целом, а также велика доля сообщений, содержащих «язык вражды». В таких сообщениях реже встречается логическое обоснование позиции, упоминание личного опыта и опыта близких. Это подтвердило гипотезу о том, что авторы негативистских, агрессивных сообщений часто находятся на «магическом» или «силовом» уровнях мышления (Пронина, 2006), хотя исследователи в целом не исключают возможность проявления агрессии и на рациональном уровне – это чаще всего проявление высокомерия, сарказм, ирония (Седов, 2005).

Было выявлено, что *агрессия в отношении оппонента* с большой вероятностью вызывала ответную агрессию, что является косвенным признаком поляризации общества. Агрессия к другим субъектам или объектам провоцировала такую же агрессию, «заражая» других пользователей подобным настроением. Сообщения с высоким уровнем *агрессии к другим объектам и субъектам* часто содержали еще и «язык вражды».

Таким образом, подтвердилась рабочая гипотеза о том, что негативно окрашенные сообщения влияют на других пользователей, создают деструктивный и травмирующий контекст обсуждения.

**Математическая модель распространения агрессии и стратегии управления**

На основе собранных данных была разработана простая математическая модель распространения агрессии в социальной сети.

За основу была взята типичная ситуация, когда в ответ на актуальный пост пользователи пишут комментарии и/или публикуют посты, содержащие или не содержащие агрессию. На следующем шаге они могут снова спровоцировать ответные комментарии, также различные по своей тональности.

В качестве единицы времени были взяты одни сутки. Именно в этот период наблюдается наиболее значимый отклик на пост после его размещения (более 90% комментариев), после активность реакции стремительно падает.

Выражение динамики агрессии оказалось следующим (использована формула полной вероятности)[11]:

$$x(t+1) = (1-\alpha)p_{АП} + \alpha[x(t)p_{АА} + (1-x(t))p_{АН}].$$

Рассмотрев предельный случай и обозначив

$$\lim_{t \to \infty} x(t) = x^*,$$

авторы получили *равновесный уровень агрессии* (долю агрессивных комментариев) – уровень, который не меняется со временем при стабильных значениях параметров:

$$x^* = \frac{(1-\alpha)p_{АП} + \alpha p_{АН}}{1 - \alpha(p_{АА} - p_{АН})}.$$

В рамках данной модели этот показатель со временем достигается при любом уровне агрессии в комментариях (и высоком, и низком).

На основании анализа основных возможностей управления в сети были выделены *два типа стратегий,* исходя из задачи снизить уровень агрессии: *«мягкая» - управление с помощью добавления комментариев* (нейтральных, позитивно окрашенных[12]) и *«жесткая» - удаление агрессивных комментариев как вариант модерации.* Первый вариант применяется, преимущественно, организациями и компаниями на независимых сайтах отзывов, хостингах, форумах. Второй характерен для официальных пабликов, сообществ, онлайн-представительств. В процессе анализа не исключалось, что оба типа стратегий могут применяться и одновременно.

Проанализируем эффективность «мягкого» и «жесткого» типов стратегий на основе данных страницы РБК в социальной сети «ВКонтакте».

Оценим начальную долю агрессивных комментариев для модели, исходя из числа агрессивных и неагрессивных комментариев в рабочей выборке из 400 комментариев. На такой же выборке посчитаем условные вероятности.

Рассмотрим два случая динамики агрессии: агрессия по отношению к оппоненту; агрессия по отношению к другим субъектам или объектам.

*Агрессия по отношению к оппоненту*. Расчет равновесного уровня агрессии дает 16% комментариев, что соответствует результатам имитационного моделирования (график синего цвета на Рис. 3).

*Рис. 3*. **Динамика сообщений, содержащих агрессию к оппоненту**

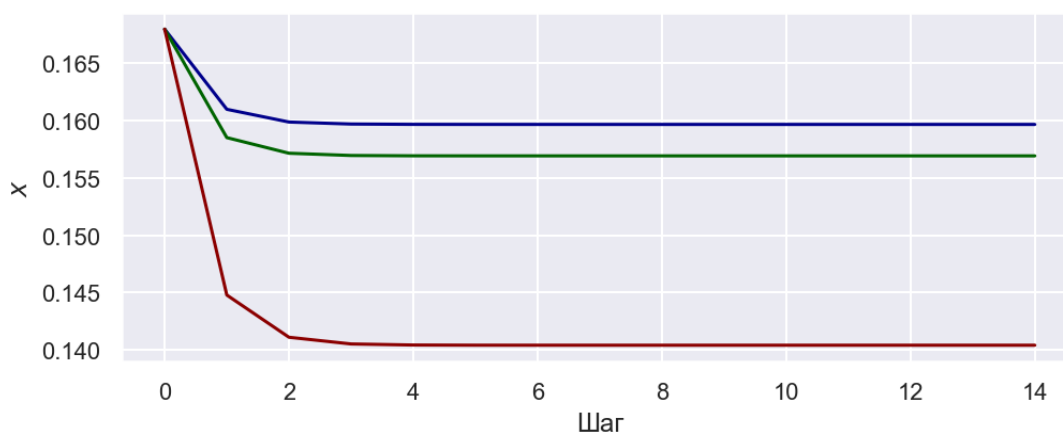

Таким образом, уровень агрессии по отношению к оппонентам с течением времени снижается даже без вмешательства «извне» в первые же сутки после публикации поста/комментария, что обусловлено снижением его актуальности. Этот процесс можно несколько ускорить при помощи управления: добавление 75 комментариев в день (10% от их общего числа) приводит к снижению доли агрессивных сообщений до 15,7% (график зеленого цвета), а удаление 75 комментариев – до 14% (график красного цвета).

В длительной перспективе применение обеих стратегий, раздельное или совместное, позволяет добиться одинакового эффекта – максимально возможного уменьшения уровня агрессии к оппоненту до 13,4%. В рамках действующей модели этот показатель снизить далее невозможно, поскольку некоторое количество агрессивных комментариев появляется на каждом шаге, даже если раньше тон обсуждения был нейтральным.

*Агрессия по отношению к другим субъектам или объектам*. Расчет равновесного уровня агрессии дает 37,3 % комментариев, имитационное моделирование приводит к такому же результату (см. Рис. 4, график синего цвета).

*Рис. 4*. **Динамика сообщений, содержащих агрессию к другим субъектам или объектам**

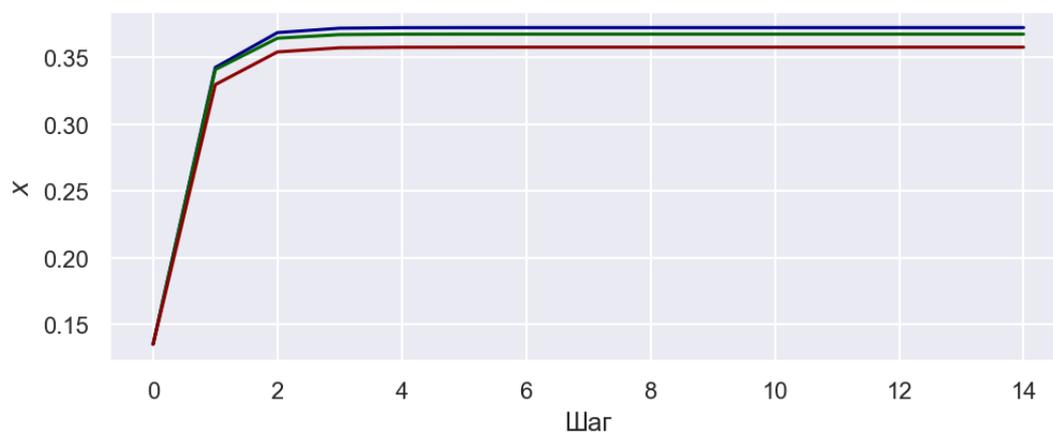

У*ровень агрессии к другим субъектам или объектам*, в отличие от уровня агрессии к собеседнику, сначала повышается, а затем остается на том же уровне, не снижаясь. Это связано с тем, что при таком типе высказываний пользователи склонны отвечать агрессивно даже на нейтральные сообщения.

Добавление 10% нейтральных/позитивных комментариев уменьшает уровень агрессии до 36,8% (график зеленого цвета), а удаление такой же доли сообщений – до 35,8% (график красного цвета). Разница в эффективности невелика, сопоставима и скорость достижения результата. Таким образом, сообщения этого рода мало поддаются управлению, независимо от стратегии.

В долгосрочной перспективе также можно добиться максимального уменьшения равновесного уровня агрессии – он составляет 32,6%.

## Выводы

В ходе исследования было выявлено, что на страницах новых медиа при обсуждении острых социальных вопросов с высокой вероятностью присутствует некий базовый (равновесный) уровень агрессии, который присутствует в ветке обсуждения постоянно, даже если предшествующие сообщения были позитивными и/или нейтральными. Это особенность социальных медиа, которую можно отнести к рискам их использования наряду с распространением фейков, буллингом, троллингом, мошенничеством, и ее важно учитывать.

В паблике РБК «ВКонтакте», например, при обсуждении тематики COVID-19 *доля сообщений с агрессией к собеседнику* составила 16%, далее этот уровень снижался только до 13,4% независимо от стратегий управления – они помогали только ускорить этот процесс. *Доля комментариев с агрессией к другим объектам* (стране, чиновникам, вакцинам и т.д.) составила 37,3%, и этот показатель также не опускался ниже 32,6%, с применением методик управления или при их отсутствии.

Разные стратегии воздействия (удаление негативных сообщений или добавление позитивных/нейтральных) показали приблизительно одинаковую эффективность. Их выбор зависит от информационной политики страницы (паблика), не исключено и применение обеих стратегий одновременно: это отвечает задачам профилей СМИ, где важно и продемонстрировать полифонию мнений, и не допустить грубого нарушения принципа информационного баланса.

**Примечания**

[1] Рагнедда М. Концептуализация и измерение цифрового капитала. // МедиаТренды. 2019. №8 (71). Режим доступа: http://www.journ.msu.ru/downloads/2019/MediaTrendi_71.pdf (дата обращения 15.05.2022).

[2] По данным Mediascope AdIndex City Conference, 2021. [Электронный ресурс]. Режим доступа:https://mediascope.net/upload/iblock/cd5/Adindex%20City%20Conference%202021%20Mediascope.pdf (дата обращения 03.03.2022).

[3] Global Didital 2022. Отчет We Are Social, Hootsuite, 2022. [Электронный ресурс]. Режим доступа: file:///C:/Users/Asus-PC/Downloads/Digital_2022_Russian_Federation.pdf (дата обращения 10.05.2022).

[4] Киберагрессия – термин, обозначающий проявления девиантного поведения в Интернете. К нему относятся оскорбления, унижения, издевательства, разоблачения, агрессивные нападки, преследования посредством коммуникативных технологий, манипулирование. Введен в 2007 году доктором философии Д. Шабро.

[5] Радулова Н. Страх и ненависть в рунете. Почему в российском сегменте Интернета столько агрессии. // Огонек. 2020. №32. С 28.

[6] Исследование платформы Perfluence, сервиса психотерапии Zigmund.Online и агентства аналитики ResearchMe. Респондентами стали пользователи от 18 до 55 лет (более 5 тыс. чел.). 2021. [Электронный ресурс]. Режим доступа: https://www.sostav.ru/publication/issledovanie-perfluence-i-zigmund-online-52273.html (дата обращения 25.03.2022).

[7] Paw Research Center. Many Americans Say Made-Up News Is a Critical Problem That Needs To Be Fixed. 05.06.2019. Access mode: Many Americans Say Made-Up News Is a Critical Problem That Needs To Be Fixed | Pew Research Center

⁸ Brand Analytics, 2022. [Электронный ресурс]. Режим доступа: https://br-analytics.ru/blog/social-aggressiveness-2021/ (дата обращения 04.04.2022).

⁹ Страница РБК ВКонтакте. [Электронный ресурс]. Режим доступа: https://vk.com/wall-25232578 (дата обращения 25.05.2022).

¹⁰ Библиотека Python Data Analysis Library. [Электронный ресурс]. Режим доступа: https://pandas.pydata.org (дата обращения 04.06.2022).

Библиотека Numerical Python. [Электронный ресурс]. Режим доступа: https://numpy.org (дата обращения 04.06.2022).

Библиотека Scientific Library for Python. [Электронный ресурс]. Режим доступа: https://scipy.org (дата обращения 04.06.2022).

¹¹ В формуле использованы следующие обозначения:

- $x(t) \in [0; 1]$ – доля агрессивных комментариев в сети на шаге $t = 1, 2, ...$,
- $\alpha$ – вероятность того, что на шаге $t$ комментарий будет написан в ответ на другой комментарий (соответственно, с вероятностью $(1 - \alpha)$ комментарий будет написан в ответ на пост),
- $p_{АП}$ – вероятность того, что комментарий будет агрессивным, если он пишется в ответ на пост,
- $p_{АА}$ – вероятность того, что комментарий будет агрессивным, если он пишется в ответ на агрессивный комментарий,
- $p_{АН}$ – вероятность того, что комментарий будет агрессивным, если он пишется в ответ на неагрессивный комментарий,
- $N$ – количество комментариев, публикуемое на каждом шаге $t$.

¹² При этом имеется в виду только тональность сообщений, а не отношение к чему-либо, например, к вакцинации.

# Библиография

**Notes**

Brand Analytics. (2022). Available at: https://br-analytics.ru/blog/social-aggressiveness-2021/ (accessed: 04.04.2022).

Global Didital (2022). We Are Social, Hootsuite. Available at: file:///C:/Users/Asus-PC/Downloads/Digital_2022_Russian_Federation.pdf (accessed: 10.05.2022).